\documentclass{article}
\input epsf.sty
\newcommand{\half}{\frac{1}{2}}
\newcommand{\bright}{\begin{flushright}}
\newcommand{\eright}{\end{flushright}}
\newcommand{\bminip}{\begin{minipage}}
\newcommand{\eminip}{\end{minipage}}
\newcommand{\bcent}{\begin{center}}
\newcommand{\ecent}{\end{center}}

\newcommand{\nnb}{\nonumber}
\newcommand{\reflef}{(\ref}
\newcommand{\MP}{M_{\rm P}}
\newcommand{\BBbox}{\mbox{\Large $\sqcap$}\hspace{-1.0em}\mbox{\Large $\sqcup$}}\newcommand{\lmd}{\lambda}
\newcommand{\Lmd}{\Lambda}
\newcommand{\gsim}{\mbox{\raisebox{-.3em}{$\;\stackrel{>}{\sim}\;$}}}
\newcommand{\lsim}{\mbox{\raisebox{-.3em}{$\;\stackrel{<}{\sim}\;$}}}

\begin{document}
\baselineskip=0.48cm
\bcent
{\Large\bf Some aspects of the scalar-tensor theory}$\footnote[2]{Based on the talks delivered at Second Advanced Research Workshop, Gravity, Astrophysics, and Strings, Kiten, Bulgaria, June 10-16, 2004}$\\[.2em]
Yasunori Fujii$\footnote[3]{E-mail: fujii@e07.itscom.net}$ \\
Advanced Research Institute for Science and Engineering,\\
Waseda University, Tokyo, 169-8555 Japan  
\ecent
\mbox{}\\[-1.9em]

\bibliographystyle{plain}
\textwidth=156mm
\textheight=239mm
\oddsidemargin=5mm
\evensidemargin=5mm
\topmargin=-10mm

\bcent
\bminip{13.5cm}
\bcent
{\bf Abstract}
\ecent
\mbox{}\\[-1.4em]
\hspace*{1em}The scalar-tensor theory of gravitation has been and still is one of the most widely discussed ``alternative theories" to General Relativity (GR).  Despite nearly half a century of its age, it continues to attract renewed interests of not only theorists but also experimentalists when we now face such issues like the accelerating universe and possible time-variability of the fine-structure constant, both viewed as something beyond the standard GR. It appears that the theory provides realistic results sometimes even beyond what is expected from the quintessence approach aimed primarily to be more phenomenological.  It seems nevertheless as if some of the unique aspects of this theory are not fully understood, even leading to occasional confusions.  I try in my lectures, partly using the contents of our book (Y.F. and K. Maeda, Scalar-tensor theory of gravitation, Cambridge University Press, 2003), to discuss some of the most crucial concepts starting from elementary introduction to the theory.  Particular emphases will be placed on the unique features of the nonminimal coupling term, the roles of the conformal transformations together with the choice of a physical conformal frame and the value of the coupling strength to the matter. Readers are advised to refer to the references for more details. 
\eminip
\ecent

\section{Basics}
\subsection{Jordan-Brans-Dicke models}

The scalar-tensor theory was invented first by P. Jordan \cite{jordan} in the 1950's, and then taken over by C. Brans and R.H. Dicke \cite{BD} some years later.  We start with the simplest and well-known models due to these pioneers.  The basic Lagrangian is given by
\begin{equation}
{\cal L}_{\rm JBD}= \sqrt{-g}\left( \varphi R - \omega\frac{1}{\varphi}g^{\mu\nu}\partial_\mu\varphi \partial_\nu\varphi +L_{\rm matter}  \right), 
\label{bsl1-1}
\end{equation}
where we use the reduced Planckian unit system in which $c=\hbar=\MP \left( =(8\pi G)^{-1/2} \right) =1$, with the units of length, time and energy given by the conventional units:
\begin{equation}
8.07 \times 10^{-33}{\rm cm},\quad 2.71\times 10^{-43}{\rm sec},\quad
2.44\times 10^{18}{\rm GeV}, \quad (13.8 {\rm Gy}\approx 10^{60.2}),
\label{bsl1-3}
\end{equation}
where the last entry for the present age of the universe is of particular interest in what follows.  We also use the notations somewhat different from the originals:
\begin{equation}
\varphi = \half\xi \phi^2, \quad \epsilon \xi^{-1}= 4\omega, \quad \epsilon = {\rm Sign} (\omega),\quad \xi >0,
\label{bsl1-2}
\end{equation}
putting \reflef{bsl1-1}) into the form
\begin{equation}
{\cal L}_{\rm JBD} = \sqrt{-g}\left(\half \xi \phi^2 R -\epsilon \half g^{\mu\nu}\partial_\mu\phi \partial_\nu\phi +L_{\rm matter}  \right).
\label{bsl1-4}
\end{equation}
The first term on the right-hand side, called "nonminimal coupling term" (NM) unique to the scalar-tensor theory, replaces the Einstein-Hilbert term (EH) in the standard theory:
\begin{equation}
{\cal L}_{\rm EH} = \sqrt{-g}\frac{1}{16\pi G}R.  
\label{bsl1-5}
\end{equation}
Comparing this with the NM term we find that this theory does not contain a truly ``constant" gravitational constant $G$, replaced by an ``effective gravitational constant" defined by
\begin{equation}
\frac{1}{8\pi G_{\rm eff}}= \xi \phi^2,
\label{bsl1-6}
\end{equation}
which is spacetime-dependent through the scalar field $\phi(x)$.  One of the pioneers' motivations was obviously to provide a theory which accommodates a time-dependent $G$ as had been suggested by P. Dirac \cite{dirac}.  His ``large-numbers hypothesis" as well as the prediction $G(t)\sim t^{-1}$ do not seem fully accepted at the present time.  Nevertheless his idea of certain fundamental physical ``constants" which may not be true constants left a tremendous impact on physicists' way of thinking in the subsequent years.  Now in the era of ``unified theories" such ``variable constants," including the fine-structure constant $\alpha$ in addition to $G$, are even considered to be the expected signs of the presence of a deeper theory behind the phenomenological world as we see it.

This view is even corroborated when string theory, supposed to be one of the most promising theoretical models of unification, contains a scalar field, often called dilaton as a spinless partner of the tensor metric field in higher dimensional spacetime, appearing with precisely the same coupling as had been shown by the scalar-tensor theory.  It appears as if the scalar-tensor theory was re-discovered 20 years later by string theory of the 1970's.

A word on the sign of the second term on the right-hand side of \reflef{bsl1-4}).  Our sign convention is such that the kinetic energy of $\phi$ is positive for $\epsilon = +1$, whereas $\epsilon = -1$ implies a ghost having a negative energy. We point out, however, that the negative energy in this sense dose not always pose an immediate difficulty, because $\phi$ is mixed with the spinless component of $g_{\mu\nu}$ through the NM term; we have to go through the process of ``diagonalization'' before we determine the sign of the energy of the diagonalized scalar field, or ``normal modes.''

One of the things to be pointed out is that Jordan admitted the scalar field to be included in the matter Lagrangian $L_{\rm matter}$, whereas Brans and Dicke (BD) assumed not, because only in this way one can save Weak Equivalence Principle (WEP).  For this reason the name ``BD model'' seems appropriate to the assumed absence of $\phi$ in $L_{\rm matter}$.  This is a theoretical model to start with, however, finally to be revised in the presence of the cosmological constant, a crucial ingredient to understand the accelerating universe.  Notice, also, that possible (small) violation of WEP is now suspected to be a generic consequence of string theory, as discussed at the end of this Section.

\subsection{Field equations}

Varying \reflef{bsl1-4}) with respect to $g^{\mu\nu}$ and $\phi$ gives the (extended) Einstein's equation and the $\phi$ equation, respectively:\\
\hspace*{9.5em}
\bminip{2cm}
\[
\left\{
\begin{array}{l}
\\
\\
\\
\end{array}
\right.
\]
\eminip
\hspace{-6em}
\bminip{12cm}
\begin{eqnarray}
2\varphi G_{\mu\nu}&=& T_{\mu\nu}+T_{\mu\nu}^{\phi}
-2\left( g_{\mu\nu}\BBbox -\nabla_{\mu}\nabla_{\nu}  \right)\varphi, \label{bsl1-7} \\
\BBbox\varphi &=& \zeta^2 T, \quad\mbox{with}\quad \zeta^{-2}=6+\epsilon\xi^{-1}=6+4\omega,  \label{bsl1-8} \\
\nabla_{\mu}T^{\mu\nu}&=& 0, \label{bsl1-9} 
\end{eqnarray}
\eminip
\mbox{}\\[.6em]
where the last equation \reflef{bsl1-9}) comes from the Bianchi identity, and 
\begin{equation}
\frac{\delta \left( \sqrt{-g}L_{\rm matter}\right)}{\delta g^{\mu\nu}}
=-\half \sqrt{-g}T_{\mu\nu},\quad
T^{\phi}_{\mu\nu}=\epsilon\left( \partial_{\mu}\phi  \partial_{\nu}\phi
	-\half g_{\mu\nu} \left( \partial\phi  \right)^2 \right).
\label{bsl1-10}
\end{equation}
The exact way to derive these equations are too complicated to be presented here.  More details will be found in our book \cite{cup}, though we outline briefly how \reflef{bsl1-8}) is obtained.

We  start with
\begin{equation}
\xi\phi R+\epsilon\BBbox\phi =0, 
\label{bsl1-11}
\end{equation}
as derived immediately from \reflef{bsl1-4}).  Multiply this with $\phi$ to obtain
\begin{equation}
2\varphi R+\epsilon\phi\BBbox\phi =0,
\label{bsl1-12}
\end{equation}
where
\begin{equation}
\BBbox\phi=\frac{1}{\sqrt{-g}}\partial_{\mu}\left( \sqrt{-g}g^{\mu\nu} \partial_{\nu}\phi \right).
\label{bsl1-13}
\end{equation}
Taking a trace of \reflef{bsl1-7}) we find
\begin{equation}
-2\varphi R=T-\epsilon \left( \partial\phi \right)^2 -6\BBbox\varphi.
\label{bsl1-14}
\end{equation}
Adding this to \reflef{bsl1-12}), and using $\BBbox\phi^2 =2\left( \phi\BBbox\phi + (\partial\phi)^2  \right)$ verified in the sense of \reflef{bsl1-13}), we finally arrive at \reflef{bsl1-8}).

From Bianchi identity of \reflef{bsl1-7}) we first derive
\begin{equation}
\nabla_{\mu}T^{\mu\nu}=-\nabla_{\mu}T^{\mu\nu}_{\phi}
+2\left(\left[\nabla^{\nu},\BBbox \right] +G^{\mu\nu}\partial_{\mu}
\right)\varphi,
\label{bsl1-15}
\end{equation}
the right-hand side turns out to vanish after some complicated algebras.  In view of the terms other than $T_{\mu\nu}$ on the right-hand side of \reflef{bsl1-7}), it may even seem accidental to obtain a simple result like \reflef{bsl1-9}), though one may understand it from the argument which is found in the discussion related to Fig. 1 in the following.

Also as is well-known, covariantized conservation law $\nabla_\mu T^{\mu\nu}=0$ applied to a point particle entails a geodesic equation:
\begin{equation}
\frac{d^2x^\mu}{d\tau^2}+\Gamma^\mu_{\ \nu\lmd}\frac{dx^\nu}{d\tau}\frac{dx^\lmd}{d\tau}=0,
\label{bsl1-16}
\end{equation}
which expresses Universal Free Fall, or WEP.  On the other hand, the latter will be understood later on the basis of the NM.

\subsection{Weak-field approximation}

Apply expansions
\begin{equation}
g_{\mu\nu}=\eta_{\mu\nu}+M_P^{-1}h_{\mu\nu}(x),\quad \mbox{and}\quad
\phi= v +Z\sigma,
\label{bsl1-17}
\end{equation}
where $\MP^{-1}$ has been re-installed to remind that the second term on the first equation is of the order of $\sim G^{1/2}$, while $v$ is the ``vacuum value."  The coefficient $Z$ is still unknown but the relevant result is independent of $Z$.  The ``weak-field" implies to maintain only the linear terms with respect to $h_{\mu\nu}$ and $\sigma$.  For $\sigma=0$, the NM term should reproduce the EH term, hence deriving
\begin{equation}
\xi v^2=1,\quad\mbox{thus}\quad v=\xi^{-1/2},
\label{bsl1-18}
\end{equation}
giving also
\begin{equation}
\varphi = \half \xi\phi^2 \approx \half +\xi^{1/2}Z\sigma.
\label{bsl1-19}
\end{equation}
Substituting this into \reflef{bsl1-8}) gives
\begin{equation}
\BBbox\sigma =\xi^{-1/2}Z^{-1}\zeta^2 T.
\label{bsl1-20}
\end{equation}
For a static point mass at the origin we have $T\approx -\rho = -M \delta(\vec{r})$, yielding
\begin{equation}
\sigma(\vec{r})=\frac{M}{M_P} \xi^{-1/2}Z^{-1}\zeta^{2}\frac{1}{4\pi r}.
\label{bsl1-21}
\end{equation}

The linear terms of \reflef{bsl1-7}) becomes
\begin{eqnarray}
\BBbox h_{\mu\nu} 
-\partial_{\mu}\partial_{\lambda}h^{\lambda}_{\nu}
-\partial_{\nu}\partial_{\lambda}h^{\lambda}_{\mu}
+\partial_{\mu}\partial_{\nu}h
&+&\eta_{\mu\nu}\left( \partial_{\rho}  \partial_{\sigma}h^{\rho\sigma}
 -\BBbox h \right) \nnb\\
&&\hspace{-.5em}-4Z\xi^{1/2}\left( \eta_{\mu\nu}\BBbox -\partial_{\mu}\partial_{\mu}  \right)\sigma = -2\MP^{-1}T_{\mu\nu}.
\label{bsl1-22}
\end{eqnarray}
The last term on the left-hand side represents a mixing coupling between $\sigma$ and the spinless part of the second-rank tensor field.  This comes from the NM coupling term, which can also be expanded:
\begin{equation}
\sqrt{-g} (\xi/2)\phi^2 R\approx \xi v \left(\partial_\mu \sigma\right) \left(  \partial_\nu h^{\nu \mu} -\partial^\mu h \right)+\cdots.
\label{bsl1-23}
\end{equation}

In the presence of mixing, a standard way is to introduce a diagonalized field $\chi_{\mu\nu}$ defined by
\begin{equation}
\chi_{\mu\nu}=h_{\mu\nu}-\half \eta_{\mu\nu}h -2 Z\xi^{1/2} \eta_{\mu\nu}\sigma, \quad\mbox{or}\quad  h_{\mu\nu}=\chi_{\mu\nu}-\half \eta_{\mu\nu}\chi -2Z \xi^{1/2}\eta_{\mu\nu}\sigma.
\label{bsl1-24}
\end{equation}
With the aid of the gauge condition $\partial_{\lambda}\chi^{\lambda}_{\nu}=0$, \reflef{bsl1-22}) is put into a simple form:
\begin{equation}
\BBbox\chi_{\mu\nu}=-\frac{2}{\MP}T_{\mu\nu},
\label{bsl1-25}
\end{equation}
from which a static solution for a point-particle source follows:
\begin{equation}
\chi_{00}(\vec{r})=2\frac{M}{\MP}\frac{1}{4\pi r},\qquad \mbox{other components}=0.
\label{bsl1-26}
\end{equation}

Due to \reflef{bsl1-16}), the theory is considered to be a ``geometrical theory," yielding the Newtonian limit with the gravitational potential for a test particle of mass $m$ given in terms of $h_{00}$, as in GR.  According to the second of \reflef{bsl1-24}) substituted from \reflef{bsl1-21}) and \reflef{bsl1-26}), we find
\begin{equation}
V=-\half \frac{m}{\MP}h_{00}=V_{\chi}+V_{\sigma},\quad\mbox{with}\quad   
V_{\chi} = -\half\frac{mM}{\MP^2}\frac{1}{4\pi r},\quad 
V_{\sigma} =  -\frac{mM}{\MP^2}\zeta^2 \frac{1}{4\pi r}.
\label{bsl1-27}
\end{equation}

We find $V_\sigma<0$ if $\zeta^2 >0$.  From a simple second-order perturbation for the one-$\sigma$-exchanged diagram, on the other hand, we know that the scalar force is attractive if $\sigma$ has a positive energy.  Combining these two arguments, we reach a conclusion that $\sigma$ is a normal field, not a ghost, if $\zeta^2 >0$.

Furthermore, $\zeta^{-2}$ as given by the second of \reflef{bsl1-8}), can be positive even if $\epsilon = -1$, as is the case in string theory or Kaluza-Klein (KK) way of deriving a scalar field. This is basically the same as the well-known situation in which the energies of the (diagonalized) normal modes are different from the starting modes of oscillation due to the (off-diagonal) mode coupling.  In the present case including a gauge field $h_{\mu\nu}$, however, the mixing coupling occurs in the kinetic energy term with derivatives, as represented explicitly in \reflef{bsl1-23}).  \\[-8.5em]


\hspace*{-1.8em}
\bminip[t]{7.4cm}
\mbox{}\\[5.em]

\hspace*{1em}To the present lowest-order perturbative approximation, the effect is represented by a diagram in Fig. 1.  Notice that the pole $k^{-2}$ coming from a graviton propagator is canceled by the derivative $k_\mu k_\nu -\eta_{\mu\nu}k^2$ coming from \reflef{bsl1-23}).  Also remember from BD's premise on the absence of $\phi$ in $L_{\rm matter}$ that the only interaction of $\sigma$ (apart from the minimal coupling) is the NM term.  It then follows that the only way for $\sigma$ to have a matter coupling is through this diagram, leading effectively to the coupling through $T$, which depends only on the total energy of the matter system.  This is the reason why the absence of a direct $\phi$-matter coupling in the Lagrangian assures WEP. 
\eminip
\hspace{1.2em}
\bminip[t]{7.7cm}
\vspace{.5em}
\baselineskip=0.4cm
\hspace*{-1.5cm}
\epsfxsize=10cm
\epsffile{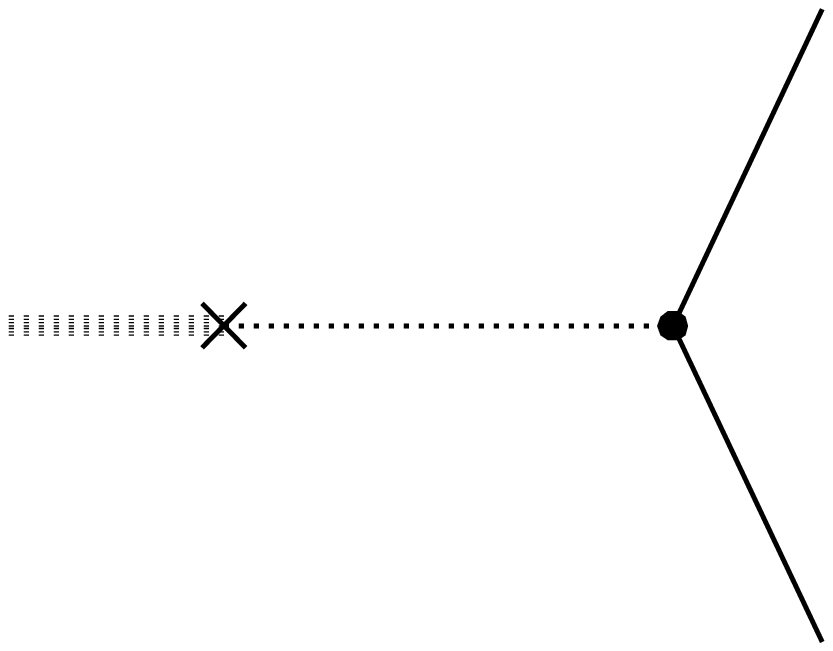}
\mbox{}\\[-7.8em]
Figure 1: The lowest-order diagram for the $\sigma$-matter coupling.  A cross represents the mixing \reflef{bsl1-23}), while a filled circle is for $T_{\mu\nu}$, the ordinary matter energy-momentum tensor for the metric-matter coupling. Heavy dotted line, dotted line and  solid line are for $\sigma, h_{\mu\nu}$, and a matter particle, respectively.
\eminip
\mbox{}\\

Let us discuss Parametrized Post Newtonian (PPN) approximation, which provides
\begin{equation}
-g_{00}\approx 1-\frac{a_g}{r} +\frac{\beta -\gamma}{2}\:\frac{a_g^2}{r^2}, \quad\mbox{and}\quad
g_{rr}\approx 1+\gamma\frac{a_g}{r}.
\label{bsl1-27a}
\end{equation}
Assuming that GR is broken only through the presence of the scalar field, we find
\begin{equation}
\beta = 1-2\zeta^2, \quad \gamma = 1-4\zeta^2.
\label{bsl1-28}
\end{equation}
Now the latest of the solar-system experiments \cite{cassini} gives the stringent result $|\:\gamma -1.0\: |\lsim 2\times 10^{-5}$, yielding the constraint:
\begin{equation}
\zeta^2\sim \xi \lsim 5\times 10^{-6},\quad\mbox{or}\quad \omega \gsim 50,000,
\label{bsl1-29}
\end{equation}
giving an {\em unnaturally} small $\xi$ as long as it is a fundamental constant of the theory.

\hspace*{-1.8em}
\bminip[t]{7.4cm}
\mbox{}\\[-2.1em]

\hspace*{1.0em}At this point it seems interesting to see how $\zeta^2$ defined by the second of \reflef{bsl1-8}), restricted to be positive, behaves as a function of $\xi$. As shown in Fig. 2, we find very different behaviors for $\epsilon = \pm 1$.  We conclude that any value of $\xi >0$ is allowed if $\epsilon = +1$, while only $\xi > 1/6$ is if $\epsilon = -1$.  Also $\zeta^2$ is restricted to be less than $1/6$ for $\epsilon = +1$.  Interesting enough, the so-called GR limit, $\zeta^2 =0$, decoupling of the scalar field from matter, is not allowed for $\epsilon = -1$, which is supported not only by such theoretical approaches like string theory and KK theory, but also by a promising cosmological model in the presence of a cosmological constant $\Lmd$, as will be shown later.
\eminip
\hspace{6.2em}
\bminip[t]{7.7cm}
\vspace{-.8em}
\baselineskip=0.4cm
\hspace*{-.5cm}
\epsfxsize=5.1cm
\epsffile{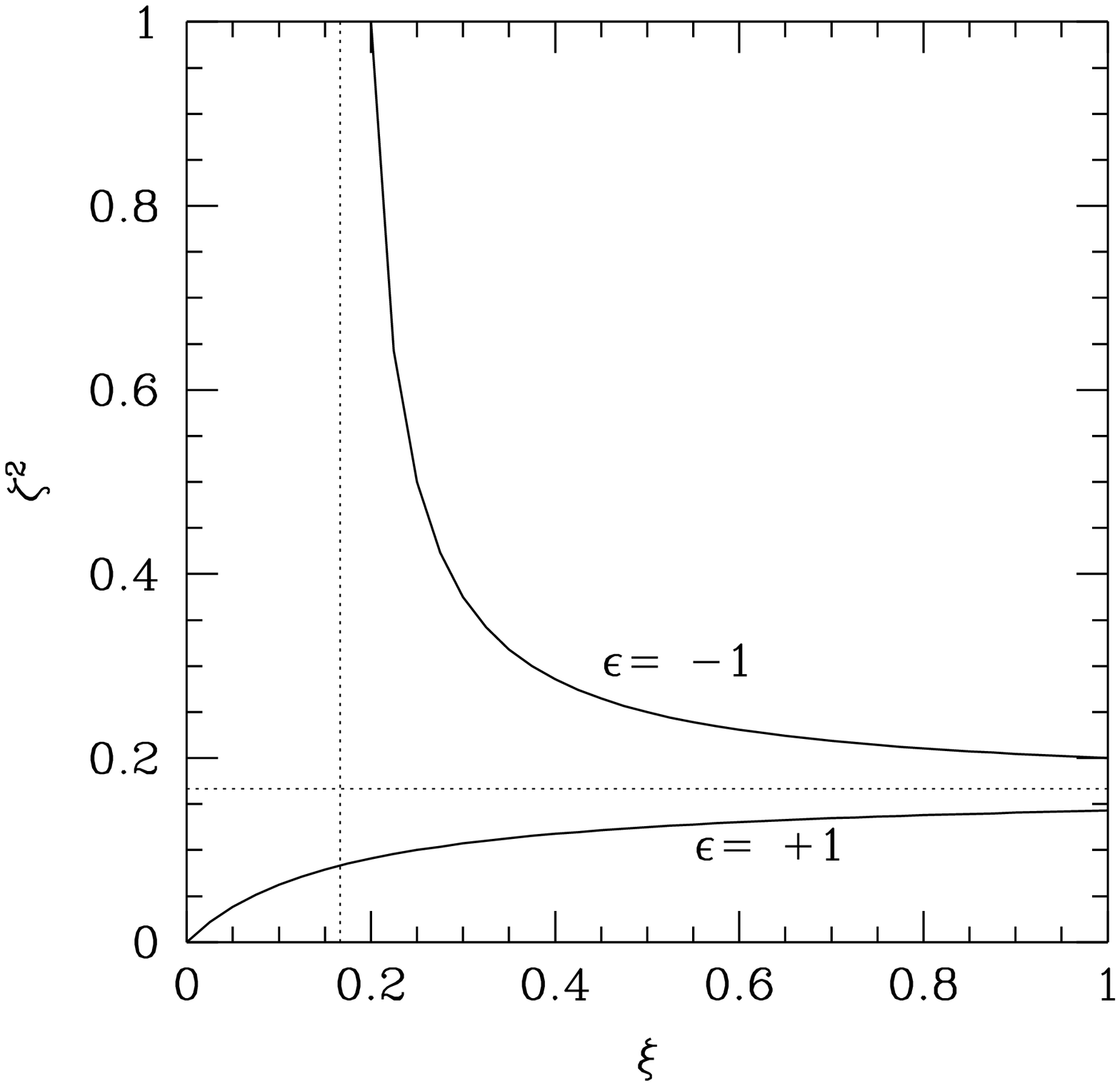}
\mbox{}\\[-.5em]
\baselineskip=0.4cm
Figure 2: $\zeta^2$ as a function of $\xi$.
\eminip
\mbox{}\\[-.2em]

One of the ways to avoid an unnaturally small $\xi$ might come from giving up the notion of a long-range scalar force.  In fact the scalar field is not a gauge field lacking immunity against acquiring a nonzero self-mass arising from the interaction with other fields in the sense of relativistic quantum field theory.  For the interaction with basically gravitational strength, the self-mass squared and the corresponding force-range are estimated, with $m_q, M_{\rm ssb}\sim {\rm TeV}$ for the quark mass and mass scale of supersymmetry breaking:
\begin{equation}
\mu_\sigma^2 \sim \frac{m_q^2 M_{\rm ssb}^2}{\MP^2}\sim (10^{-9} {\rm eV})^2,\quad\mbox{and}\quad \mu_\sigma^{-1} \sim 100{\rm m},
\label{bsl1-29a}
\end{equation}
with the latitude of several orders of magnitude. The potential $V_\sigma$ in the last term of \reflef{bsl1-27}) will then be multiplied by $e^{-r\mu_\sigma}$, hence giving a non-Newtonian potential \cite{nn}, though the coefficient will be modified according to the way described in the Subsection 3.4.  Since $\mu_\sigma^{-1}$ is so short compared with any of the sensible distances of the solar system that it seems unlikely that the solar-system experiments constrains $\zeta$, thus allowing much larger and natural value of $\xi$.

\subsection{Relation with string theory}

Strongly in this connection, we briefly discuss how string theory in higher-dimensions  provides a promising origin of the scalar field.  In the low-energy limit, a closed string separates into the metric tensor field with a spinless companion, dilaton, $\Phi$, and the antisymmetric second-rank tensor field $B_{\bar{\mu}\bar{\nu}}$ with its field strength $H_{\bar{\mu}\bar{\nu}\bar{\lambda}}$.  The relevant part of the Lagrangian is given by
\begin{equation}
{\cal L}_{\rm st}= \sqrt{-g}e^{-2\Phi}\left(
\half R +2 g^{\bar{\mu}\bar{\nu}}\partial_{\bar{\mu}} \Phi
\partial_{\bar\nu} \Phi 
 -\frac{1}{12} H_{\bar{\mu}\bar{\nu}\bar{\lambda}}
H^{\bar{\mu}\bar{\nu}\bar{\lambda}}
\right).
\label{bsl1-33}
\end{equation}
By introducing $\phi$ by $\phi =2e^{-\Phi}$, we may put the first two terms on the right-hand side of \reflef{bsl1-33}) into
\begin{equation}
{\cal L}_{\rm st1}= \sqrt{-g}\left( \half \xi \phi^2 R - \half \epsilon 
g^{\bar{\mu}\bar{\nu}}\partial_{\bar{\mu}}\phi \partial_{\bar{\nu}}\phi 
 \right),
\label{bsl1-34}
\end{equation}
which, surprisingly, shows the same appearance as the first two terms on the right-hand side of \reflef{bsl1-4}), with the identification $\epsilon =-1, \xi^{-1}=4$, hence $\omega=-1$.  
We see how naturally a ``large" value of $\xi$ arises, though \reflef{bsl1-34}) is given in higher dimensions.  We may simply expect that $\epsilon =-1$ as well as the condition $\zeta^{-2}>0$  survive dimensional reduction to 4 dimensions.  This seems to suggest an alternative approach  to the Principle of Least Coupling \cite{dmpl}, also by accepting a massive $\sigma$ field in the context discussed above.

The Lagrangian with other fields for ordinary matter fields included, however, has no protection against the matter part from being ``contaminated'' by $\phi$, thus violating BD's premise.  From this point of view, string theory appears to allow WEP violation, in general, as emphasized also by \cite{dm}, though its effect might be suppressed if the force has finite range.

\section{Conformal transformation}
\subsection{Scale transformation (Dilatation)}

Let us start with a global scale transformation, or sometimes called dilatation:
\begin{equation}
x^\mu \rightarrow x^\mu_* = \lmd x^\mu,\quad\mbox{with constant } \lmd.
\label{bsl1-41}
\end{equation}
This coordinate transformation corresponds to a uniform and (4-dimensionally) isotropic expansion or contraction in flat Minkowski spacetime. We obviously have
\begin{equation}
x^\mu = \lmd^{-1}x^\mu_*,\quad\mbox{and}\quad \partial_x = \lmd \partial_{*x}.
\label{bsl1-42}
\end{equation}

If we have only massless fields or particles, we have no way to provide a fixed length scale, and we feel no difference even if all the coordinates are uniformly dilated in accordance with \reflef{bsl1-41}).  We then have a scale invariance or dilatation symmetry. Suppose, on the other hand, we have a fundamental field or particle having a nonzero mass $m$.  The inverse $m^{-1}$ will provide a fixed length or time standard, which breaks the above-mentioned invariance.

To implement this idea, let us introduce a real free massive scalar field $\Phi$ (not to be confused with the dilaton as discussed in Section 1.4), as a representative of matter fields:  
\begin{equation}
 L_{\rm matter}= -\half (\partial \Phi)^2 -\half m^2\Phi^2, \quad (\partial \Phi)^2 \equiv \eta^{\mu\nu}(\partial_\mu\Phi)(\partial_\nu\Phi).
\label{bsl1-43z}
\end{equation}
We then find
\begin{eqnarray}
I_{\rm matter}=\int d^4 xL_{\rm matter} 
&=& \int d^4 x_* \lmd^{-4}\left( -\half \lmd^2  (\partial_* \Phi)^2 -\half m^2\Phi^2\right) \nnb\\ 
&=& \int d^4x_* \left( -\half (\partial_* \Phi_*)^2 -\half m^2\lmd^{-2}\Phi_*^2 \right),\quad\mbox{with}\quad  \Phi_* =\lmd^{-1}\Phi.
\label{bsl1-43}
\end{eqnarray}
Notice that we defined $\Phi_*$ primarily to leave the kinetic term form invariant except for putting the $*$ symbol everywhere.  In this way we guarantee the invariance for a massless field.  On the other hand, the mass term in the last equation depends on $\lmd$, thus breaking scale invariance.

As a next step, we try to extend the argument to the curved spacetime.  The concept of a global coordinate $x^\mu$ is no longer useful.  The transformation \reflef{bsl1-41}) will be replaced by the  similar transformation of the metric tensor, as will be shown again in the simplified model of a scalar field $\Phi$, but with certain modifications in the first line of \reflef{bsl1-43}):
\begin{equation}
{\cal L}_{\rm matter}= \sqrt{-g}\left( -\half (\partial \Phi)^2 -\half m^2\Phi^2 \right), \quad (\partial \Phi)^2 \equiv g^{\mu\nu}(\partial_\mu\Phi)(\partial_\nu\Phi). 
\label{bsl1-44}
\end{equation}
In place of \reflef{bsl1-41}), we may try
\begin{equation}
g_{\mu\nu}\rightarrow  g_{*\mu\nu}=\lmd^2 g_{\mu\nu},\quad\mbox{or}\quad g_{\mu\nu}=\lmd^{-2} g_{*\mu\nu},
\label{bsl1-45}
\end{equation}
from which follow also
\begin{equation}
 g^{\mu\nu}=\lmd^{2} g^{*\mu\nu},\quad \mbox{and}\quad \sqrt{-g}= \lmd^{-4}\sqrt{-g_*}.
\label{bsl1-46}
\end{equation}
This is not a general coordinate transformation, leaving the infinitesimal coordinate difference $dx^\mu$ unchanged.  We then find
\begin{eqnarray}
{\cal L}_{\rm matter}&=&  \lmd^{-4}\sqrt{-g_*}\left(-\half \lmd^2(\partial \Phi)^2 -\half m^2\Phi^2 \right) \nnb\\
&=& \sqrt{-g_*}\left(-\half (\partial_* \Phi_*)^2 -\half \lmd^{-2}m^2\Phi_*^2 \right),\quad \mbox{with}\quad \Phi_*=\lmd^{-1}\Phi.
\label{bsl1-47}
\end{eqnarray}
We introduced $\Phi_*$ with the same relation as in \reflef{bsl1-43}), finding the same appearance as in the last lines.

The global transformation discussed above, also to be called scale transformation, will turn out to be useful in developing realistic theory of scalar-tensor theory.

\subsection{Conformal transformation (Weyl rescaling)}

The global scale transformation in curved spacetime as discussed above
may be promoted to a local transformation by replacing the constant
parameter $\lmd$ by a local function $\Omega(x)$, an arbitrary function of $x$.  This defines a conformal transformation, or sometimes called Weyl rescaling:
\begin{equation}
 g_{\mu\nu}\rightarrow g_{*\mu\nu}=\Omega^{2}(x) g_{\mu\nu},\quad \mbox{or}\quad ds^2 \rightarrow ds^2_* = \Omega^{2}(x)ds^2.
\label{bsl1-51}
\end{equation}
According to the last equation, we are considering a local change of units, not a coordinate transformation.  However, the condition for invariance is somewhat more complicated than the global predecessors, as will be shown by the examples:
\begin{itemize}
\item Defining the new scalar field by $\Phi =\Omega\Phi_*$ no longer entails a complete invariance even with $m=0$, because $\partial_\mu \Phi = \Omega\left( \partial_\mu +f_\mu \right)\Phi_*$ with $f_\mu = \partial_\mu\ln \Omega$.
\item The Lagrangian ${\cal L}=-(1/4)\sqrt{-g} g^{\mu\rho}g^{\nu\sigma}F_{\mu\nu}F_{\rho\sigma}$ is left invariant with the electromagnetic potentials unchanged; $A_{*\mu}=A_\mu$ with $F_{\mu\nu}=\partial_\mu A_\nu -\partial_\nu A_\mu$.
\item Massless fermion maintains invariance only if there is no torsion.
\end{itemize}

There is no conformally invariant theoretical model of gravity of practical importance.  Let us see how the scalar-tensor theory is affected by the conformal transformation.  We start with
\begin{equation}
\partial_{\mu}g_{\nu\lambda}=\partial_{\mu}\left( \Omega^{-2}g_{*\nu\lambda} \right)=\Omega^{-2}\partial_{\mu}g_{*\nu\lambda} -2\Omega^{-3}\partial_{\mu}\Omega g_{*\nu\lambda}=\Omega^{-2}\left( \partial_{\mu}g_{*\nu\lambda}
-2f_{\mu}g_{*\nu\lambda} \right),
\label{bsl1-52}
\end{equation}
where $f=\ln\Omega, f_\mu =\partial_\mu f, f_*^\mu = g_*^{\mu\nu}f_\nu$. We then compute
\begin{equation}
\Gamma^{\mu}_{\hspace{.3em}\nu\lambda}=
\half g^{\mu\rho}\left(   
	\partial_{\nu} g_{\rho\lambda}+\partial_{\lambda} g_{\rho\nu}
	-\partial_{\rho} g_{\nu\lambda} \right)
=\Gamma^{\mu}_{*\hspace{.1em}\nu\lambda}-\left(   f_{\nu}\delta^{\mu}_{\lambda}
  +f_{\lambda}\delta^{\mu}_{\nu}-f_{*}^{\mu}g_{*\nu\lambda}
\right),
\label{bsl1-53}
\end{equation}
reaching finally
\begin{equation}
R=\Omega^2\left( R_{*}+6\BBbox_{*}f -
6 g_{*}^{\mu\nu}f_{\mu}f_{\nu}\right),
\label{bsl1-54}
\end{equation}
which indicates complications in gravity theory.

Using this in the first term on the right-hand side of \reflef{bsl1-4}) with $F(\phi) =\xi\phi^2$, we obtain
\begin{equation}
{\cal L}_1=\sqrt{-g} \half F(\phi) R =\sqrt{-g_{*}}\half F(\phi)\Omega^{-2}
\left( R_{*}+6\BBbox_{*}f -
6 g_{*}^{\mu\nu}f_{\mu}f_{\nu}\right). 
\label{bsl1-55}
\end{equation}
We may choose
\begin{equation}
F\Omega^{-2}=1, 
\label{bsl1-55a}
\end{equation}
so that the first term on the right-hand side goes to the standard EH term.  We say that we have moved to the Einstein conformal frame (E frame). We have
\begin{equation}
\Omega=F^{1/2},\quad\mbox{then } f= \ln\Omega,\hspace{1em}
f_{\mu}=\partial_\mu f=\frac{\partial_\mu\Omega}{\Omega}=\half \frac{\partial_{\mu}F}{F} =\half\frac{F'}{F}\partial_{\mu}\phi, 
\label{bsl1-56}
\end{equation}
where $F'\equiv dF/d\phi$.
The second term on the right-hand side of \reflef{bsl1-55}) then goes away by partial integration, while the third term becomes $-\sqrt{-g_*}(3/4)(F'/F)^2g_*^{\mu\nu}\partial_\mu\phi \partial_\nu\phi$.  This term is added to the second term on the right-hand side of \reflef{bsl1-4}) giving the kinetic term of $\phi$:
\begin{equation}
-\half\sqrt{-g_{*}}\Delta g^{\mu\nu}_{*}\partial_{\mu}\phi\partial_{\nu}\phi, \quad\mbox{with}\quad \Delta= \frac{3}{2}\left( \frac{F'}{F}  \right)^2
+\epsilon\frac{1}{F}.
\label{bsl1-57}
\end{equation}

If $\Delta >0$, we define a new field $\sigma$ by
\begin{equation}
\frac{d\sigma}{d\phi}= \sqrt{\Delta},\quad\mbox{hence}\quad \sqrt{\Delta}\partial_\mu\phi =\frac{d\sigma}{d\phi} \partial_\mu\phi = \partial_\mu\sigma,
\label{bsl1-58}
\end{equation}
thus bringing \reflef{bsl1-57}) to a canonical form $-(1/2)\sqrt{-g_{*}}g^{\mu\nu}_{*}\partial_{\mu}\sigma\partial_{\nu}\sigma$.  If $\Delta <0$, the opposite sign in the first expression of \reflef{bsl1-57})  propagates to the sign of the preceding expression, implying a ghost, which is excluded in what follows.

By using the explicit expression of $F(\phi)$ we find
\begin{equation}
\Delta = \left( 6+\epsilon\xi^{-1} \right)\phi^{-2} =\zeta^{-2}\phi^{-2},
\label{bsl1-59}
\end{equation}
which translates the condition $\Delta >0$ into $\zeta^2 >0$.  We further obtain
\begin{equation}
\frac{d\sigma}{d\phi}=\zeta^{-1}\phi^{-1},\quad\mbox{hence}\quad \zeta\sigma = \ln \left(\frac{\phi}{\phi_{0}} \right),\quad\mbox{or}\quad
 \phi=\xi^{-1/2}e^{\zeta\sigma},
\label{bsl1-60}
\end{equation}
reaching also
\begin{equation}
\Omega =e^{\zeta\sigma}.
\label{bsl1-60a}
\end{equation}
We finally obtain
\begin{equation}
{\cal L}_{\rm JBD}=\sqrt{-g_{*}}\left( \half R_{*} - \half
	g^{\mu\nu}_{*}\partial_{\mu}\sigma\partial_{\nu}\sigma 
	+L_{\rm *{\rm matter}}  \right).
\label{bsl1-61}
\end{equation}

Now which conformal frame did we come from?  We name the starting conformal frame the Brans-Dicke conformal frame (BD frame), instead of the Jordan frame as often called simply to imply any non-Einstein frame.  In this way we have moved, as shown schematically:\\
\begin{picture}(300,90)(-50,10)
\put(5,60){from}
\put(50,40){\framebox(110,45)}
\put(60, 70){BD conformal frame}
\put(65, 60){$G$ varies}
\put(65, 50){$m$ constant}
\put(180,60){to}
\put(210,40){\framebox(100,45)}
\put(220,70){E conformal frame}
\put(225,60){$G$ constant}
\put(225, 50){$m$ varies?}
\end{picture}
\mbox{}\\[-2.0em]

But then which is the physical conformal frame?  The answer depends crucially on the presence of the cosmological constant, as will be discussed in the next Section.

In closing this Subsection, we re-emphasize that a conformal transformation from any conformal frame to the E frame, without the mixing term, includes the process of diagonalization.

\section{$\Lambda$ cosmology}
\subsection{$\Lambda$ cosmology in the Brans-Dicke frame}

We now include $-\Lmd$ in \reflef{bsl1-4}):
\begin{equation}
{\cal L}_{{\rm BD}\Lambda}=\sqrt{-g}\left( \half \xi\phi^2 R -
	\half\epsilon g^{\mu\nu}\partial_{\mu}\phi\partial_{\nu}\phi -
	\Lambda +L_{\rm matter} \right),
\label{bsl1-66}
\end{equation}
from which follow the field equations:
\begin{eqnarray}
2\varphi G_{\mu\nu}&=& T_{\mu\nu}+T_{\mu\nu}^{\phi}-g_{\mu\nu}\Lambda -
2\left( g_{\mu\nu}\BBbox -\nabla_{\mu}\nabla_{\nu} \right)\varphi, \label{bsl1-67}\\
\BBbox\varphi &=& \zeta^2 \left( T -4\Lambda \right), \label{bsl1-68}\\
\nabla_{\mu}T^{\mu\nu}&=& 0.  \label{bsl1-69}
\end{eqnarray}
By assuming the radiation-dominated spatially flat Friedmann cosmology, \reflef{bsl1-67})-\reflef{bsl1-69}) are simplified to
\begin{eqnarray}
6\varphi H^2&=& \epsilon\half \dot{\phi}^2 +\Lambda +\rho -6
H\dot{\varphi}, \label{bsl1-70} \\
\ddot{\varphi}+3H\dot{\varphi}&=&4\zeta^2  \Lambda, \label{bsl1-71}\\
\dot{\rho} +4 H{\rho}&=&0.  \label{bsl1-72}
\end{eqnarray}

We find a surprisingly simple solution by putting 
\begin{equation}
H=0,
\label{bsl1-73}
\end{equation}
which is desperately away from ``standard cosmology,'' though this is the place where the effect of $\Lambda$ shows up most dramatically. By using this result, \reflef{bsl1-71})-\reflef{bsl1-72}) are transformed into
\begin{eqnarray}
\ddot{\varphi}&=&4\zeta^2 \Lambda,  \label{bsl1-74}\\
\dot{\rho} &=&0, \label{bsl1-75}
\end{eqnarray}
which solve to give
\begin{eqnarray}
\varphi &\approx& 2\zeta^2 \Lambda t^2,\quad \mbox{or}\quad \phi\approx \sqrt{4\Lambda \zeta^2 \xi^{-1}}t= \sqrt{\frac{4\Lmd}{6\xi+\epsilon}}t, \label{bsl1-76} \\
\rho &=& {\rm const}. \label{bsl1-77}
\end{eqnarray}
Substituting them further into \reflef{bsl1-70}) yields
\begin{equation}
\rho =-3\Lambda \frac{2\xi+\epsilon}{6\xi +\epsilon}.
\label{bsl1-78}
\end{equation}
Demanding $\rho>0$, we obtain the condition
\begin{equation}
\epsilon = -1,\quad\mbox{and}\quad 2<\xi^{-1}<6, \mbox{ hence }\zeta^2 >1/4.
\label{bsl1-79}
\end{equation}
It seems interesting to point out that $\epsilon =-1$ agrees with what string theory and KK approach suggest.\\

\hspace*{-1.8em}
\bminip[t]{7.6cm}
\mbox{}\\[-2.1em]
\hspace*{1em}
The solution shown above, $H=0$, is independent of any initial condition, so must be an {\em attractor} or asymptotic solution.  We try to find how asymptotic the solution is.  An example of numerical solutions is shown in Fig. 3, demonstrating it likely that the asymptotic state is reached rather quickly.

\hspace*{1.5em}This model is basically the same as what A.D. Dolgov tried to show; the constant energy density $\Lambda$ is canceled by the negative kinetic energy of $\phi$ \cite{dolgov}.  The idea was criticized against the simultaneous decrease of the effective gravitational ``constant,'' $G_{\rm eff}\sim \varphi^{-1}\sim t^{-2} \rightarrow 0$; a trivial world of no gravitation \cite{wbrg}. See, however, the next Subsection for the ultimate solution of this problem.

\hspace*{1.5em}
The result \reflef{bsl1-73}) disqualifies the BD frame to be acceptable as a physical conformal frame.  Then how about the E frame?
\eminip
\hspace{2.2em}
\bminip[t]{6.5cm}
\vspace{-.8em}
\baselineskip=0.4cm
\hspace*{1.5cm}
\mbox{}\\[-.5em]
\hspace{3em}
\epsfxsize=4.95cm
\epsffile{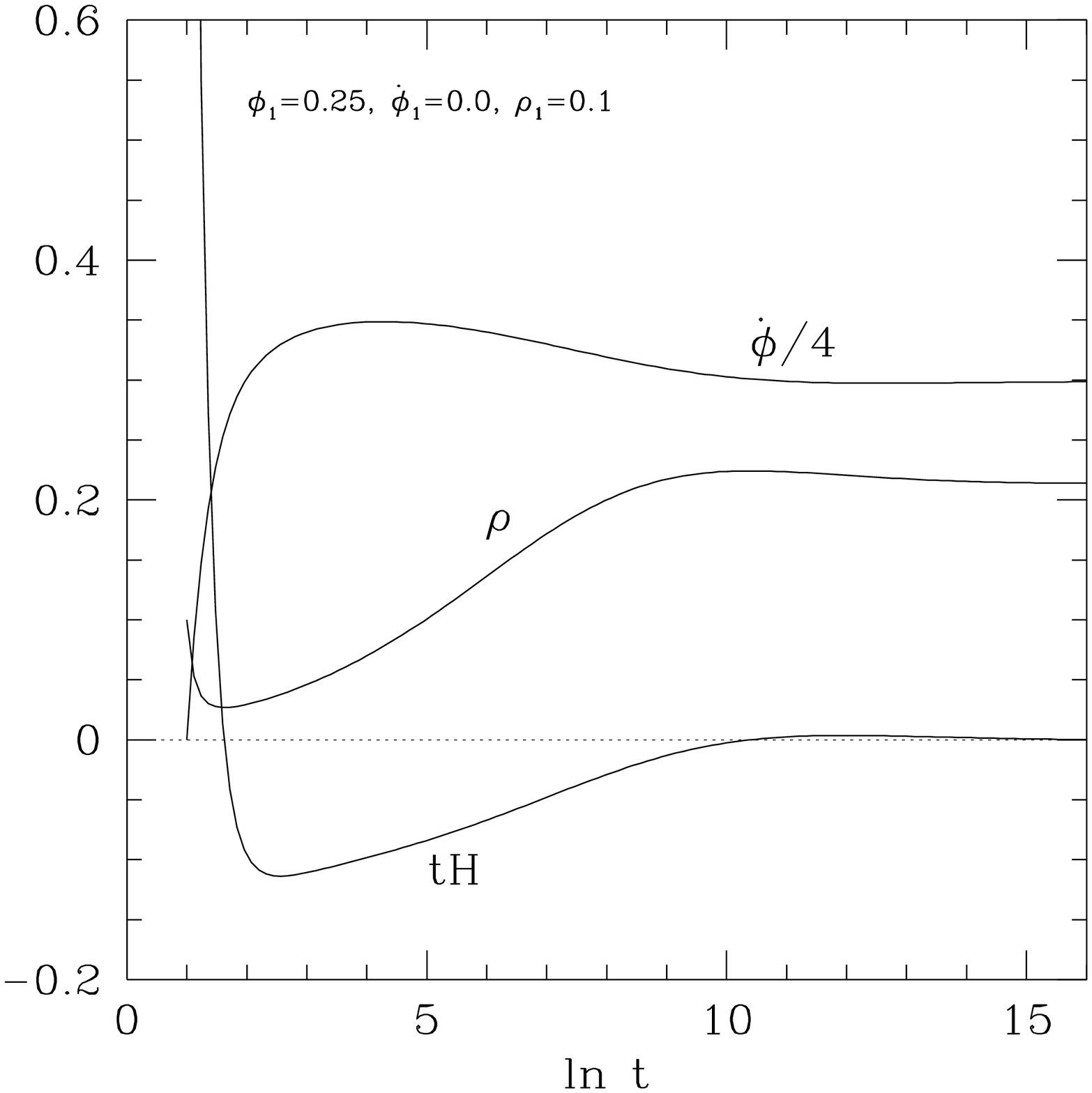}
\mbox{}\\[.5em]
Figure 3: An example of the solutions in the BD conformal frame. The asymptotic behavior $H=0$ is reached well before $\ln t_0 \sim (1/2)\ln t_{*0}\sim 70$ (cf. \reflef{bsl1-86})) for the present epoch, $\log t_{*0}\approx 60$. 
\eminip
\mbox{}\\[-.5em]

\subsection{$\Lambda$ cosmology in the Einstein frame}

Applying \reflef{bsl1-51}) with $F=\xi\phi^2$ to \reflef{bsl1-66}) yields
\begin{equation}
{\cal L}_{\rm BD\Lmd}=\sqrt{-g_{*}}\left( \half R_{*} - \half
	g^{\mu\nu}_{*}\partial_{\mu}\sigma\partial_{\nu}\sigma -V(\sigma)
	+L_{\rm *{\rm matter}}  \right),\quad\mbox{with}\quad V(\sigma)=\Omega^{-4}\Lambda = \Lmd e^{-4\zeta\sigma}. 
\label{bsl1-80}
\end{equation}
Note that this exponential potential, favored in many of the quintessence models \cite{quint} chosen, however, in an {\em ad hoc} manner, has been {\em derived} simply by conformally transforming the term $\sqrt{-g}\Lambda$.

The field equations are also obtained:
\begin{eqnarray}
&&G_{*\mu\nu}=T_{*\mu\nu}^{(\sigma)}+T_{*\mu\nu} \equiv {\cal
T}_{\mu\nu},\label{bsl1-82}\\ 
&&\BBbox_{*}\sigma +4\zeta\Lambda
e^{-4\zeta\sigma}=\zeta T_{*}, \label{bsl1-83}
\end{eqnarray}
where $T_{*\mu\nu}$ is the matter energy-momentum tensor in the E frame, while ${\cal T}_{\mu\nu}$ is defined by the far-right-hand side of \reflef{bsl1-82}).  From the obvious Bianchi identity $\nabla_\mu {\cal T}^{\mu\nu} = 0$ follows $\nabla_\mu T_*^{\mu\nu} = -g^{\nu\mu}\left( \partial_\mu \sigma\right) \zeta T_*$.  An apparently non-conservation of $T_{*\mu\nu}$ by no means results in WEP violation because the non-vanishing right-hand side is independent of any individual characteristics of matter objects.

In the spatially flat Friedmann universe, the second of \reflef{bsl1-51}) is put into the form:
\begin{equation}
-\Omega^2 dt^2 +\Omega^2 a^2(t)d\vec{x}^2= -dt_{*}^2 +a^2_{*}(t_{*})d\vec{x}^2,
\label{bsl1-84}
\end{equation}
from which follows
\begin{equation}
dt_{*}=\Omega dt,  \quad\mbox{and}\quad a_{*}=\Omega a.
\label{bsl1-85}
\end{equation}
Using \reflef{bsl1-60a}) combined with \reflef{bsl1-76}), we find $\Omega \sim t$, to be further substituted into the first of \reflef{bsl1-85}) yielding
\begin{equation}
t_* =t^2,
\label{bsl1-86}
\end{equation}
apart from inessential coefficients.  Further using the same equations we reach
 the result
\begin{equation}
\sigma (t_*)= \half \zeta^{-1}\ln t_* +\mbox{const},\quad\mbox{and}\quad a_*(t_*)= t_*^{1/2},
\label{bsl1-87}
\end{equation}
where we used the obvious result $a=\mbox{const}$, another expression of \reflef{bsl1-73}).  The second equation shows that the attractor solution in the E frame is now fully in agreement with the standard cosmology.

The $00$-th component of \reflef{bsl1-82}) gives
\begin{equation}
3H_*^2 =\rho_\sigma +\rho_*,\quad\mbox{with}\quad \rho_{\sigma}=\half \dot{\sigma}^2 +V(\sigma).
\label{bsl1-88}
\end{equation}
Using \reflef{bsl1-87}) in this equation we obtain
\begin{equation}
\Lmd_{\rm eff}=\rho_\sigma =\frac{3}{16}\zeta^{-2}t_*^{-2},\quad\mbox{and}
\quad \rho_* = \frac{3}{4}\left( 1-\frac{1}{4}\zeta^{-2} \right)t_*^{-2}.
\label{bsl1-89}
\end{equation} 
The first equation shows that the effective cosmological constant defined by $\rho_\sigma$ falls off like $t_*^{-2}$, in agreement with the {\em scenario of a decaying cosmological constant}.  If we are allowed to interpret the E frame as the physical conformal frame, and $t_*$ as a physical cosmic time, with the present age $t_{*0}\approx 1.38\times 10^{10}{\rm y} \approx 10^{60.2}$ in units of the reduced Planckian time, we find today's value to be $\Lambda_{\rm eff}\sim 10^{-120}$.  This allows the simplest way to understand so small a value without recourse to any fine-tuning of the parameters \cite{yflmd}.  This seems to be a major success of the scalar-tensor theory, at least as the first step, though further details have to be worked out to fully understand the observed accelerating universe, as elaborated in \cite{yftn}, for example.  Also in connection with the critical comment made toward the end of the preceding Subsection, we re-emphasize that $G$ is now constant.

Notice also that the condition $\rho_* >0$ requires the same result as
\reflef{bsl1-79}).  Appreciating this conclusion seriously, we decide to prefer the ``string-inspired choice," $\epsilon =-1$, together with the idea of a
massive scalar field to be free from the solar-system constraint which supports  $\epsilon= +1$.  Assuming a nonzero 
mass of $\sigma$, however, might interfere with the cosmological solution as in \reflef{bsl1-89}), a consequence of an exponentially decreasing potential given  in \reflef{bsl1-80}).  The required dual nature, the  globally (cosmologically) massless and locally massive $\sigma$, is the subject of a detailed study in \cite{yfptp}.

However, this conformal frame suffers from too much time-dependence of particle masses, as will be shown.  Consider again a real free massive scalar field $\Phi$ with the matter Lagrangian in the BD frame \reflef{bsl1-44}).  It is crucially important to note that $m$ is a pure constant because no $\phi$ is allowed according to BD's premise. Apply a conformal transformation \reflef{bsl1-51}) to move to the E frame:
\begin{equation}
{\cal L}_{\rm matter}=\sqrt{-g_*}\left( -\half g^{\mu\nu}_*\left( {\cal D}_\mu\Phi_* \right) \left({\cal D} _\nu\Phi_* \right) -\half m_*^2\Phi_*^2\right),
\label{bsl1-91}
\end{equation}
where $ {\cal D}_\mu\Phi_* =(\partial_\mu +(\partial_\mu \ln\Omega))\Phi_*$, and
\begin{equation}
m_* = \Omega^{-1}m \sim t_*^{-1/2}m,
\label{bsl1-92}
\end{equation}
where we used $\Omega \sim t$ and \reflef{bsl1-86}).

Now the physical conformal frame should be the one in which we perform astronomical observations using atomic clocks with the time standard provided by masses of microscopic particles, particularly the electron mass, taking, however, suspected slight time-dependence of the fine-structure constant aside for the moment.  The situation remains the same when we measure atomic spectra to determine the redshifts of distant objects.  We also recognize that we have absolutely no way to detect any change of the standard itself.  We thus conclude that in the physical conformal frame the particle masses should stay time-independent, in contradiction with \reflef{bsl1-92}).  We suspect the BD model to be blamed.

\subsection{Revised model for particle masses}

Consider the mass term in the E frame:
\begin{equation}
{\cal L}_{\rm mass}=-\half\sqrt{-g_*}m_\natural^2\Phi_*^2,
\label{bsl1-93}
\end{equation}
where $m_\natural$ is {\em assumed} to be constant, unlike $m_*$ in \reflef{bsl1-91}).  Try to {\em go back} to the starting conformal frame by using the same $\Omega$ as defined in \reflef{bsl1-60a}) and \reflef{bsl1-60}). The result turns out to be simple:
\begin{equation}
{\cal L}_{\rm mass}=-\half\sqrt{-g}f_\Phi^2 \phi^2\Phi^2,\quad\mbox{with}\quad f_\Phi = \xi^{1/2}\frac{m_\natural}{\MP}.  
\label{bsl1-94}
\end{equation}
This is quite different from the BD model with the mass term in \reflef{bsl1-90}).  The absence of dimensional constants suggests an invariance under the global scale transformation:
\begin{equation}
g_{\mu\nu}\rightarrow g'_{\mu\nu}=\lmd^2g_{\mu\nu}, \ \phi\rightarrow \phi'=\lmd^{-1}\phi, \ \Phi\rightarrow \Phi'=\lmd^{-1}\Phi.
\label{bsl1-95}
\end{equation}
For this reason we call the proposed model with the mass term replaced by \reflef{bsl1-94}) a scale invariant (sc) model.  In fact the description exclusively in terms of dimensionless constants can be found in all of the terms in the Lagrangian except for the $\Lambda$ term:
\begin{equation}
{\cal L}_{\rm sc\Lmd}=\sqrt{-g}\left( \half \xi\phi^2 R -
	\half\epsilon g^{\mu\nu}\partial_{\mu}\phi\partial_{\nu}\phi -
	\Lambda -\half g^{\mu\nu}\partial_\mu\Phi \partial_\nu\Phi  -\half f_\Phi^2 \phi^2\Phi^2 \right).
\label{bsl1-96}
\end{equation}
We naturally use the terminology the scale-invariant conformal frame, with the {\em partial} invariance under \reflef{bsl1-95}).  In this way we have now established the E frame qualified to be accepted as a physical conformal frame.  The decoupling of $\sigma$ from matter, however, is true only in the classical sense, as will be shown.

\subsection{Quantum effects}

Consider interactions among matter fields.  The ensuing quantum effects
will be described in the language of relativistic quantum field
theory in tangential Minkowski spacetime. Notice that we are not
ambitious enough to quantize the metric tensor field, or spacetime
itself.  As usual, we have divergences which may be dealt with by such
mathematical tools, like dimensional regularization technique, in which
we consider $D$-dimensional spacetime keeping $D$ off the physical value
4 until the end of the calculation.  Then loops will develop infinities
as expressed by poles, particularly $(D-4)^{-1}$ at the 1-loop level.
Correspondingly, we reformulate the preceding equations into those in $D$-dimensional curved spacetime.  In some occasions we have explicit dependence on $D-4$.  In fact \reflef{bsl1-93}) is modified by multiplication of the factor
\begin{equation}
\Omega^{(D-4)}=e^{(D-4)\zeta\sigma}= 1+(D-4)\zeta\sigma + \cdots.
\label{bsl1-97}
\end{equation}
\mbox{}\\[-2.6em]

\hspace*{-1.76em}
\bminip[t]{7.cm}
\baselineskip=0.48cm
\mbox{}\\[-.9em]
\hspace*{1.em}
If we go to the limit $D\rightarrow 4$ at this level, then those $\sigma$-matter couplings go away, back to the classical limit. However, we are going to include such loop diagrams as shown in Fig. 4. The pole $(D-4)^{-1}$ will cancel the above factor $(D-4)$ in the linear term in \reflef{bsl1-97}), leaving us with a nonzero finite value for the $\sigma$-matter coupling.  This is precisely the way many examples of the so-called {\em quantum anomaly} have been calculated, effectively violating symmetries established at a classical level.
\eminip
\hspace{1.7em}
\bminip[t]{7.4cm}
\baselineskip=0.4em
\epsfxsize=8cm
\mbox{}\\[-6.8em]
\hspace*{-2.em}
\epsffile{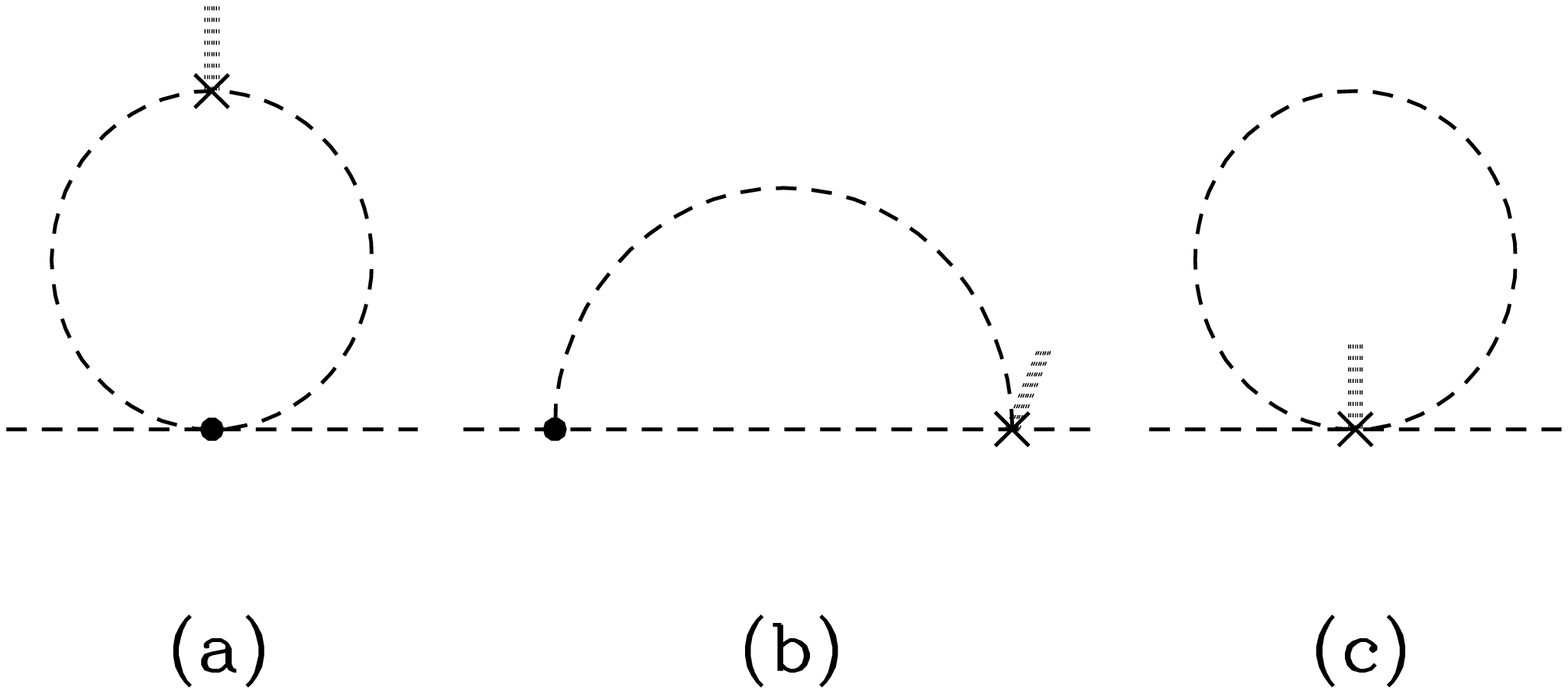}
\mbox{}\\[-3.1em]
Figure 4: Examples of 1-loop diagrams for the interaction $\lmd_\Phi \Phi_*^4$, represented, together with the derived 3-vertex, by a filled circle.  Heavy dotted lines are for $\sigma$.
\eminip
\mbox{}\\[-.0 em]

We then go into many details and complications, including such subjects as re-adjusting the physical conformal frame, $\dot{G}/G$, WEP violation, non-Newtonian force, fate of broken scale invariance, among other things.  At the present time we still do not know exactly what the destination of this approach is going to be like. We still emphasize that the scalar-tensor theory, compared with the more phenomenology-oriented quintessence approach, has provided and is going to provide better understandings of many issues related to the cosmological constant, as one may find in our book \cite{cup}, at least partially.


\bigskip

\end{document}